\documentclass{Interspeech2024}
\usepackage{multirow}
\usepackage[table,xcdraw]{xcolor}
\usepackage{graphicx}
\usepackage{subfigure}

\interspeechcameraready

\title{On the Effectiveness of Acoustic BPE in Decoder-Only TTS}

\name[affiliation={1}]{Bohan}{Li}

\name[affiliation={1}]{Feiyu}{Shen}

\name[affiliation={1}]{Yiwei}{Guo}

\name[affiliation={2}]{Shuai}{Wang}

\name[affiliation={1}]{Xie}{Chen}

\name[affiliation={1, ^\ast}]{Kai}{Yu}

\address{
  $^1$MoE Key Lab of Artificial Intelligence, AI Institute, X-LANCE Lab,
Department of Computer Science and Engineering, Shanghai Jiao Tong University, China\\
  $^2$Shenzhen Research Institute of Big Data, CUHK-Shenzhen, China 
  }
\email{
  \{everlastingnight, francis\_sfy, yiwei.guo, chenxie95, kai.yu\}@sjtu.edu.cn$^1$ \\
  \{wangshuai\}@cuhk.edu.cn$^2$
}

\keywords{discrete speech token, acoustic byte-pair encoding, decoder-only text-to-speech}

\begin{document}

\maketitle

\begin{abstract}
    
    Discretizing speech into tokens and generating them by a decoder-only model have been a promising direction for text-to-speech (TTS) and spoken language modeling (SLM). To shorten the sequence length of speech tokens, acoustic byte-pair encoding (BPE) has emerged in SLM that treats speech tokens from self-supervised semantic representations as characters to further compress the token sequence. But the gain in TTS has not been fully investigated, and the proper choice of acoustic BPE remains unclear. In this work, we conduct a comprehensive study on various settings of acoustic BPE to explore its effectiveness in decoder-only TTS models with semantic speech tokens. Experiments on LibriTTS verify that acoustic BPE uniformly increases the intelligibility and diversity of synthesized speech, while showing different features across BPE settings. Hence, acoustic BPE is a favorable tool for decoder-only TTS.
\end{abstract}

\section{Introduction}

\footnotetext{$\ast$: The corresponding author.}
The language modeling paradigm has been revolutionizing text-to-speech (TTS) by its strong generation ability since the birth of large decoder-only language models (LMs). There have been LM-based TTS models that exhibits versatile, expressive and even emergent abilities \cite{GSLM, AudioLM, VALLE, SpearTTS, zhu2023vectok, BaseTTS}. 

Unlike text, speech is intrinsically a continuous signal with a much lower information density. To adapt speech into LMs and generate it in an autoregressive manner, researchers have adopted different methods to map speech into discrete tokens~\cite{AudioLM,SpearTTS}. By the purpose of discretization, discrete speech tokens can be divided into acoustic tokens~\cite{SoundStream,encodec,DAC} that aim to reconstruct the signal perfectly, and semantic tokens~\cite{vq-wav2vec,hubert,w2vBERT,WavLM}, from self-supervised models that provide a compact abstraction of speech semantics.
After discretization, decoder-only LMs like VALL-E~\cite{VALLE} can be seamlessly applied to TTS, where the discrete speech tokens are treated as the targets given a text input.

However, the low rate of information behind speech still leads to excessively lengthy discrete speech sequences compared to text transcriptions~\cite{shen2024acoustic}. For example, more than 500 HuBERT~\cite{hubert} tokens may be required for a single sentence of around 30 words just to convey an idea or two. It is even worse for the acoustic tokens, since most of neural speech codec models require a short frame shift and the residual vector quantization~\cite{encodec} technique to reconstruct the signal decently. Regardless of the type of tokens, this feature of speech poses a great challenge for long-context modeling of LM-based TTS systems. 

To address this issue, one possible way is to further compress the discrete speech sequence. 
A promising approach is the acoustic byte-pair encoding (BPE) technique, which is proposed in \cite{ren2022speech}. It is a similar method to the traditional BPE algorithm~\cite{bert} in natural language processing. It treats the discrete indexes of speech as literal characters and iteratively compresses consecutive tokens based on the frequency in the training corpus. Such compression will coherently reduce sequence length with the increase of vocabulary size. For speech discrete tokens, usually a group of multiple tokens occur together to represent a specific phoneme or syllable, and organizing them to be a unique modeling unit would provide a higher abstraction of semantics and morphological information. Therefore, it is intuitively reasonable to apply acoustic BPE on discrete speech tokens in both reducing sequence length and improving representability. In previous researches, such acoustic BPE has been adopted encode pseudo-target label in HuBERT pretraining~\cite{ren2022speech} and automatic speech recognition~\cite{chang2023exploration}.

Nevertheless, the effectiveness of applying acoustic BPE in the TTS task still remains unclear to the literature. Although BASE-TTS~\cite{BaseTTS} and VoxtLM ~\cite{VoxtLM} mentions the use of acoustic BPE in generation, the design space of acoustic BPE is still not fully investigated, and the gain of such technique in TTS needs to be further explored. Despite of the possible improvements, acoustic BPE could bring more difficulties in choosing the correct unit for generation, since the vocabulary of the LM in TTS is greatly enlarged. Too much abstraction of acoustic units could also make language modeling harder. Moreover, acoustic BPE might have different behaviors and performances on different types of discrete speech tokens. 

Therefore, in this work, we conduct a comprehensive study on the effectiveness of introducing acoustic BPE to TTS in the decoder-only LM paradigm. We implement various settings of acoustic BPE on the semantic tokens extracted by speech self-supervised models, and then train a VALL-E autoregressive decoder-only transformer as the acoustic model to generate acoustic BPEs from text inputs. Afterwards, the original semantic tokens are unfolded and fed to a unit-based vocoder~\cite{du2022vqtts, du2024unicats} for waveform synthesis. We consider HuBERT~\cite{hubert} and WavLM~\cite{WavLM} as the source of semantic tokens, adjust the number of clusters in extracting the semantic tokens from 2048 to 8192, and increase the vocabulary size of acoustic BPE up to 20k in order to observe the effects made by different acoustic BPE settings. We perform most of the experiments on LibriTTS~\cite{LibriTTS} and evaluate the effectiveness of such settings via speech intelligibility, sample diversity and speech quality in objective and subjective measurements. 

Our findings suggest that models employing acoustic BPE method are capable of generating audio with high intelligibility and quality, while also achieving much faster training and inference processes. This indicates that acoustic BPE can be of great value for the TTS task when properly configured. This could lay a solid foundation for future decoder-only LM-based TTS models, as the pros of acoustic BPE weighs much more than its con. 

\section{Decoder-Only TTS with Acoustic BPE}

\subsection{Acoustic BPE}
\label{acoustic bpe}

BPE is a widely-used algorithm for data compression and text encoding. In natural language processing,  it is initialized with a vocabulary that contains all the characters with their emergence frequency in the text corpus, and  iteratively merges the most frequent character pairs until a specified number of merges or a target vocabulary size is reached~\cite{BPE}. For most languages like English, there exists blanks as the obvious boundaries between words in the text. However, the textual sequences of audio discrete tokens lack obvious boundaries, akin to linguistic features found in Chinese. Following \cite{shen2024acoustic}, we first obtain speech discrete tokens from k-means clusters of features derived from speech self-supervised learning (SSL) models (HuBERT, WavLM). Then we establish a bijective mapping for conversion between the speech discrete tokens and the Chinese-character Unicodes, simply using an integer offset. Chinese characters mapping to speech discrete tokens are used to train the BPE model. With these operations, the acoustic BPE encoding is composed of conversion to Chinese characters and the encoding of pretrained BPE model.  The acoustic BPE decoding is exactly the inverse process. This research utilizes the acoustic BPE method as a part of the tokenizer, which plays a role in preprocessing inputs for the decoder-only language model mentioned in the next subsection(\ref{decoder-only TTS}). Similar to natural language processing, we can consider the upcoming modeling process as a spoken language modeling process.

Based on the method description above, it is evident that the method involves multiple dimensions of configuration, such as the SSL model, k-means cluster number, BPE vocabulary size, and so on. Configuring combinations across different dimensions may lead to entirely new characteristics, which can have significant implications in spoken language modeling. A comprehensive and systematic exploration of these configurations' impact on modeling is crucial for better constructing spoken language models.


\subsection{ Decoder-Only TTS Language Modeling}
\label{decoder-only TTS}

The language model can estimate the probability of the input sequence, which is usually expressed as:{
\setlength\abovedisplayskip{0cm}
\setlength\belowdisplayskip{0cm}
\begin{equation} 
p(\mathbf{x}|\theta)=\prod_{i=1}^t p(x_i | x_{<i}, \theta)
\label{equatation1} 
\end{equation}
}where $\mathbf{x}=\{x_1,...,x_t\}$ is the input sequence with length $t$ and the language model is parameterized by $\theta$.
%

However, the spoken language model in TTS pays the other attention on text conditions. Based on the decoder-only Transformer architecture, the training and inference process of the TTS model follows the autoregressive (AR) stage in VALL-E.


\subsubsection{Decoder-only TTS model}

During training process, we put the phoneme sequence $\mathbf{x}$ and tokenized audio sequence $\mathbf{s}$  into the decoder-only language model (with parameters denoted as $\theta$) , estimating the probability autoregressively, formulated as:{
\setlength\abovedisplayskip{0pt}
\setlength\belowdisplayskip{0pt}
\begin{equation} 
p(\mathbf{s}| \mathbf{x}; \theta) = \prod_{i=1}^t p(\mathbf{s}_{i}| \mathbf{s}_{:i-1},\mathbf{x}; \theta) 
\label{equatation2} 
\end{equation}
}

%
The model optimizes its parameters by performing the task of predicting the next acoustic feature based on the phonemized text and historical speech features. This training process can be viewed as maximizing the sequence probability of the tokenized audio sequence under the condition of the text, which can be described as:{
\setlength\abovedisplayskip{0pt}
\setlength\belowdisplayskip{0pt}
\begin{equation} 
\hat{\theta} = \arg\max\limits_{\theta} p(\mathbf{s}|\mathbf{x};\theta) =\arg\max\limits_{\theta}\prod_{i=1}^t p(\mathbf{s}_{i}| \mathbf{s}_{:i-1} 
,\mathbf{x};\theta)
\label{equatation3}
\end{equation}
}where $\hat{\theta}$  is our optimization target.


\subsubsection{Inference with prompts}
During inference, the model generates speech with similar speakers and prosody as the given audio prompt. Concatenating tokenized audio prompt $\mathbf{s}_\text{prompt}$ , phonemized source text $\mathbf{x}$ , and the phonemized text prompt $\mathbf{x}_\text{prompt}$ corresponding to the audio prompt, we build the input of the spoken language model in format of $\{\mathbf{x}_\text{prompt}, \mathbf{x}, \mathbf{s}_\text{prompt}\}$ . The model then autoregressively generates the remaining sequence after this sequence. The decoding process consists of acoustic BPE decoding (if used) and the vocoding process of a discrete-unit-based vocoder~\cite{du2024unicats} for speech waveform synthesis. Additionally, the Mel-spectrogram extracted from the speech prompt is also input into the vocoder to achieve better speaker control.

\section{Experiments and Results}

In this section, we will introduce the exploration involving the configuration of various dimensions in the acoustic BPE method and analyze the results regarding Decoder-only TTS performance with these configurations.
\subsection{Experimental Setup}
\subsubsection{Datasets}

The experiments were conducted on the LibriSpeech~\cite{librispeech} and LibriTTS~\cite{LibriTTS} datasets. LibriTTS, designed for text-to-speech tasks, consists of approximately 585 hours of English speech from multiple speakers. Its train-960 subset was used for model training. The LibriSpeech dataset served primarily as the test set for speech synthesis. In this experiment, sentences shorter than 4 seconds or longer than 10 seconds were filtered out from the test-clean subset of LibriSpeech. The remaining 1145 sentences, totaling approximately 2.02 hours of speech, were used as test cases. All speech audio data was downsampled to 16kHz before use.

\subsubsection{Settings of acoustic BPE}

The encoding process of acoustic BPE consists of two stages: speech discretization and acoustic BPE model training. In the speech discretization stage, we utilized the HuBERT-large\footnote{https://github.com/facebookresearch/fairseq/tree/main/examples/hubert} model pretrained with masked prediction on the 60k hours of LibriLight~\cite{librilight} dataset and the WavLM-large\footnote{https://huggingface.co/microsoft/wavlm-large} model pretrained on extra dataset consisting of 10k hours of Gigaspeech~\cite{gigaspeech} and 24k hours English data subset of VoxPopuli~\cite{voxpopuli}, as the self-supervised speech representation model. Continuous audio features were extracted from the final layer's output activations of their Transformer encoder, to train k-means models with 256, 2048, 4096 and 8192 centroids, which encoded the semantic features extracted from the LibriTTS datasets into speech discrete token sequences. Due to memory constraints, we randomly sampled 100 hours of speech from the LibriTTS train-960 set during k-means model training. 

In the next stage, we trained the acoustic BPE model on the speech discrete token sequences extracted from the LibriTTS train-960 set. All discrete token sequences were first converted into Unicode strings, and then a BPE model was training on these strings with the SentencePiece\footnote{https://github.com/google/sentencepiece} toolkit. The pretrained BPE model encoded the speech discrete token sequences from LibriTTS into acoustic BPE sequences. Four different acoustic BPE encodings were experimented with in this study: no acoustic BPE encoding, and acoustic BPE encoding with vocabulary sizes of 5,000, 10,000, and 20,000 subwords, respectively.

\subsubsection{Decoder-only TTS architecture and training}

The TTS model is based on the AR model of VALL-E~\cite{VALLE}, which is a decoder-only Transformer architecture consisting of 12 Transformer layers, each with 16 attention heads and a hidden feature dimension of 1024. 
The absolute positional encoding based on trigonometric functions is employed separately to the text and speech feature sequences.
As there are no official implementations, we resorted to an unofficial repository\footnote{https://github.com/lifeiteng/vall-e}.

During training, the model was optimized using the ScaledAdam~\cite{yao2024zipformer} optimizer and the Eden~\cite{yao2024zipformer} learning rate scheduler with an initial learning rate of 0.05. 
Each batch of training data contained approximately 100 seconds of audio data, including only speech samples with duration between 1 second and 15 seconds. To achieve a larger effective batch size, gradients were accumulated four times before each update. The model was trained for 20 epochs on an NVIDIA A10 GPU.

\subsubsection{Vocoder setup}
In each setting of the SSL extractor and k-means clustering, we trained a discrete-unit-based vocoder to convert semantic tokens into waveform\footnote{https://github.com/X-LANCE/UniCATS-CTX-vec2wav}. This vocoder contains two conformer blocks each with 2 layers and 184 attention dimensions. After the conformer blocks, a HifiGAN~\cite{hifigan} is cascaded and the same learning criterions apply.
We trained all the replicas for 1M steps on LibriTTS, and shared them within different acoustic BPE settings.

\subsubsection{Evaluation metrics}
To conduct a thorough investigation on the effectiveness of the acoustic BPE, we established the evaluation process based on a range of metrics, encompassing both subjective and objective dimensions. 
Contrasting with scenarios where this method is absent, these metrics serve to provide an intuitive way of gauging the impact of acoustic BPE. 
Below is an explanation of the metrics utilized.

\noindent\textbf{Speech Intelligibility: }We used a publicly available automatic speech recognition (ASR) model based on the Conformer-Transducer architecture\footnote{https://huggingface.co/nvidia/stt\_en\_conformer\_transducer\_xlarge} to transcribe synthesized speech into text, and then computed the word error rate (WER) between the transcribed text and the ground truth text.

\noindent\textbf{Speech Quality and Naturalness: }We employed naturalness Mean Option Score (MOS) as the subjective metric of speech quality and naturalness. In the listening test, each participant was assigned with multiple test cases, each containing high-intelligibility speech synthesized by different settings from the same sentence. Participants were asked to give a score of 1-5 to each speech based on its naturalness. Mel cepstral distortion (MCD)~\cite{mcd} with dynamic time warping (DTW), measuring the MFCC distance between the synthesized and reference mel-spectrum features with DTW helping to find the optimal time-sequence alignment, was utilized as the objective metrics.

\noindent\textbf{Inference Speed: }This research focuses on the inference speed of the decoder-only language model, which is the main architecture of the TTS system. Real-time factor (RTF) was measured to evaluate the inference speed.

\noindent\textbf{Sample Diversity: }Number of statistically-different bins (NDB) ~\cite{richardson2018gans} and Jensen-Shannon (JS) divergence  were employed to objectively evaluate the sample diversity. Detailed calculating process will be introduced in Section \ref{ndb_js}.
\begin{table}[]
\centering
\vspace{-1em}
\setlength{\abovecaptionskip}{0.1cm}%
\caption{Results of the decoder-only TTS model performance (WER, MOS) and inference speed (RTF) metrics under different BPE vocabulary sizes of HuBERT+kmeans (kms) 2048 centroids. Here ``resyn." means resynthesis from vocoder, while ``aBPE" refers to the BPE vocabulary size.}
\label{Table1}
\setlength{\tabcolsep}{4pt}
\begin{tabular}{cccccc}
\hline
\textbf{Tokenizer}                                                      & \textbf{Task}               & \textbf{aBPE}       & \textbf{WER$\downarrow$}           & \textbf{MOS$\uparrow$}           & \textbf{RTF$\downarrow$}           \\ \hline
    & resyn.                         & -                        & 1.9                         &   4.39 $\pm$ 0.13                      & -                         \\ \cline{2-6}
    &                             & -                        & 9.1                         &  4.13 $\pm$ 0.19                       & 0.129                         \\
    &                             & 5000                       & 5.7                         &   \textbf{4.20} $\pm$ 0.17                      & 0.069                         \\
    &                             & 10000                      & 5.5                         &   4.19 $\pm$ {0.13}                      & 0.052                         \\
\multirow{-5}{*}{\begin{tabular}[c]{@{}c@{}}HuBERT\\ kms 2048\end{tabular}} & \multirow{-4}{*}{TTS}       & 20000                      & \textbf{5.2}                         &   4.11 $\pm$ 0.19                      & \textbf{0.045}                         \\ \hline
\end{tabular}
\vspace{-1.7em}
\end{table}

\begin{table*}[ht!]
\centering
\setlength{\abovecaptionskip}{0.1cm}%
\caption{The objective metrics (WER and MCD) of synthetic speech under different SSL, k-means and acoustic BPE settings. Results are presented in the format: [w/o. BPE] / [w. BPE 10000]. }
\label{Table3}
\begin{tabular}{ccccccccc}
\hline
\textbf{SSL}     & \multicolumn{4}{c}{\textbf{HuBERT}}                          & \multicolumn{4}{c}{\textbf{WavLM}}                           \\ \hline
\textbf{K-means} & \textbf{256} & \textbf{2048} & \textbf{4096} & \textbf{8192} & \textbf{256} & \textbf{2048} & \textbf{4096} & \textbf{8192} \\ \hline
WER $\downarrow$             & 6.0 / 6.3        & 9.1 / 5.5       & 7.7 / \textbf{4.5}       & 7.8 / 8.0       & 9.9 / 6.0        & 8.3 / 7.8       & 11.4 / 9.0      & 9.0 / 9.5       \\
MCD $\downarrow$             & 11.9 / 12.6  & 12.1 / 12.0   & 11.9 / 11.8   & 11.9 / \textbf{11.7}   & 11.9 / 12.5  & 12.2 / 12.4    & 12.3 / 12.3   & 11.9 / 12.0   \\ \hline
\end{tabular}
\vspace{-1.6em}
\end{table*}

\subsection{Acoustic BPE Enhances TTS Model Performance}

After our preliminary experiments, it was found that in general situation, incorporating the acoustic BPE method indeed leads to significant improvements in the TTS model's ability to synthesize speech. We primarily focus on intelligibility and quality of synthesized speech to reflect this ability. The results of selected outperforming settings are presented in Table \ref{Table1}.

\subsubsection{Improvement of speech integibility and quality}
We conducted integibility experiments on the entire test set, which consists of 1145 sentences. The WER of the synthesized audios ASR results are calculated. As shown in Table \ref{Table1}, the intelligibility performance of  audio generated by TTS models using acoustic BPE method has a significant advantage over those without using it. Under the pre-configuration (tokenizer composed of HuBERT-large and a 2048-centroid kmeans model), the reduction is up to 3.9\% WER. The results of the MOS metric indicate that models using the acoustic BPE method can generate competitive audio quality. In the worst-case scenario, it is slightly lower by 0.02 compared to not using it, and it is accompanied by lower confidence. However, in the best-case scenario, it can be higher by 0.07, accompanied by better confidence.

\subsubsection{Acceleration of inference and training process}
The calculation of RTF in this research is equivalently defined as the ratio of the docoder-only LM inference time to the length of its input. As the presented result, it significantly decreases as the expansion of its vocabulary increase. This is due to the merging operations of BPE, which shorten the sequence length of the input language model. The resulting acceleration effect of the model far exceeds the time spent on the increased computational cost of token embeddings due to the expansion of the vocabulary. From an intuitive perspective, it results in up to 2.9 times inference speedup. Additionally, we record the time cost of the whole training process, the settings with acoustic BPE method also perform approximately 1.5 $\sim$ 2 times acceleration decided to configurations.

\subsection{Enrichment of sample diversity}
We are interested in investigating whether the acoustic BPE method can influence the generation of more diverse intonations or speech rates when reading the same sentence. 
We opted to utilize NDB and JS divergence as metrics for assessing the diversity of samples in this TTS model. Experimental results from various configurations are presented in Table \ref{Table2}.
\subsubsection{The NDB and JS divergence metrics}
\label{ndb_js}
The NDB metric is proposed in ~\cite{richardson2018gans}.
In our experiment, we extracted prosodic features\footnote{Namely Kaldi-style~\cite{kaldi} pitch, probability of voice and energy.} from selected low-WER synthesized utterances in the test set. Additionally, with speaker information already present in prompts, the influence of semantic and speaker information on sample diversity can be almost negligible. Samples, considered as the prosodic features of frames, were divided into a training and evaluation set. We first set a fixed number $k$ of cluster bins by training a k-means model with samples in the training set, maintaining $k$ centroids. Let $p$ and $q$ denote the statistical distributions of samples in the training and evaluation set on these bins, and let $m$ and $n$ denote the sample numbers of the training and evaluation set, respectively. The calculation of NDB metric 
was conducted with a \textit{two-proportion z-test} on $p$, $q$ and $m$, $n$, counting the number of the bins whose final \textit{p}-values are less than a manually set value of \textit{significant level}, and dividing it by $k$. 

The definition of Jensen-Shannon (JS) divergence is 
{
\setlength\abovedisplayskip{0.2cm}
\setlength\belowdisplayskip{0.2cm}
\begin{equation}
JS = \frac{1}{2}\left[KLD\left(p \mid\mid \frac{p+q}{2}\right) + KLD\left(q \mid\mid \frac{p+q}{2}\right)\right]
\end{equation}
}where $KLD(\cdot || \cdot)$ is the Kullback–Leibler (KL) divergence function. To reduce the impact of uncertain selection of two sample sets, we calculated two metrics using a public tool\footnote{https://github.com/eitanrich/gans-n-gmms/blob/master/utils/ndb.py} for 10 times and used their average values as the final results. Greater values of NDB and JS indicate a larger difference between the statistical distributions of the samples, suggesting better performance in terms of diversity.

\begin{table}[]
\centering
\setlength{\abovecaptionskip}{0.1cm}%
\caption{Sample diversity results. Here ``kms" means kmeans centroids, and ``aBPE" refers to the BPE vocabulary size.}
\label{Table2}
\setlength{\abovecaptionskip}{0cm}
\vspace{0em} 
\begin{tabular}{cccc}
\hline
\textbf{Tokenizer}           & \textbf{aBPE} & \textbf{NDB$\uparrow$} & \textbf{JS$\uparrow$} \\ \hline
\multirow{4}{*}{HuBERT kms 2048} & -                  & 0.649             & 0.00313            \\
                             & 5000               & 0.676             & \textbf{0.00439}   \\
                             & 10000              & 0.668             & 0.00396            \\
                             & 20000              & \textbf{0.678}    & 0.00388            \\ \hline
\multirow{4}{*}{WavLM kms 2048}  & -                  & 0.665             & 0.00364            \\
                             & 5000               & 0.680             & 0.00415            \\
                             & 10000              & \textbf{0.687}    & 0.00408            \\
                             & 20000              & 0.684             & \textbf{0.00453}   \\ \hline
\end{tabular}
\vspace{-2.0em}
\end{table}
\begin{figure}[htbp]
    \centering
    \setlength{\abovecaptionskip}{0.cm}
    \subfigure[NDB]{\includegraphics[width=1.5in]{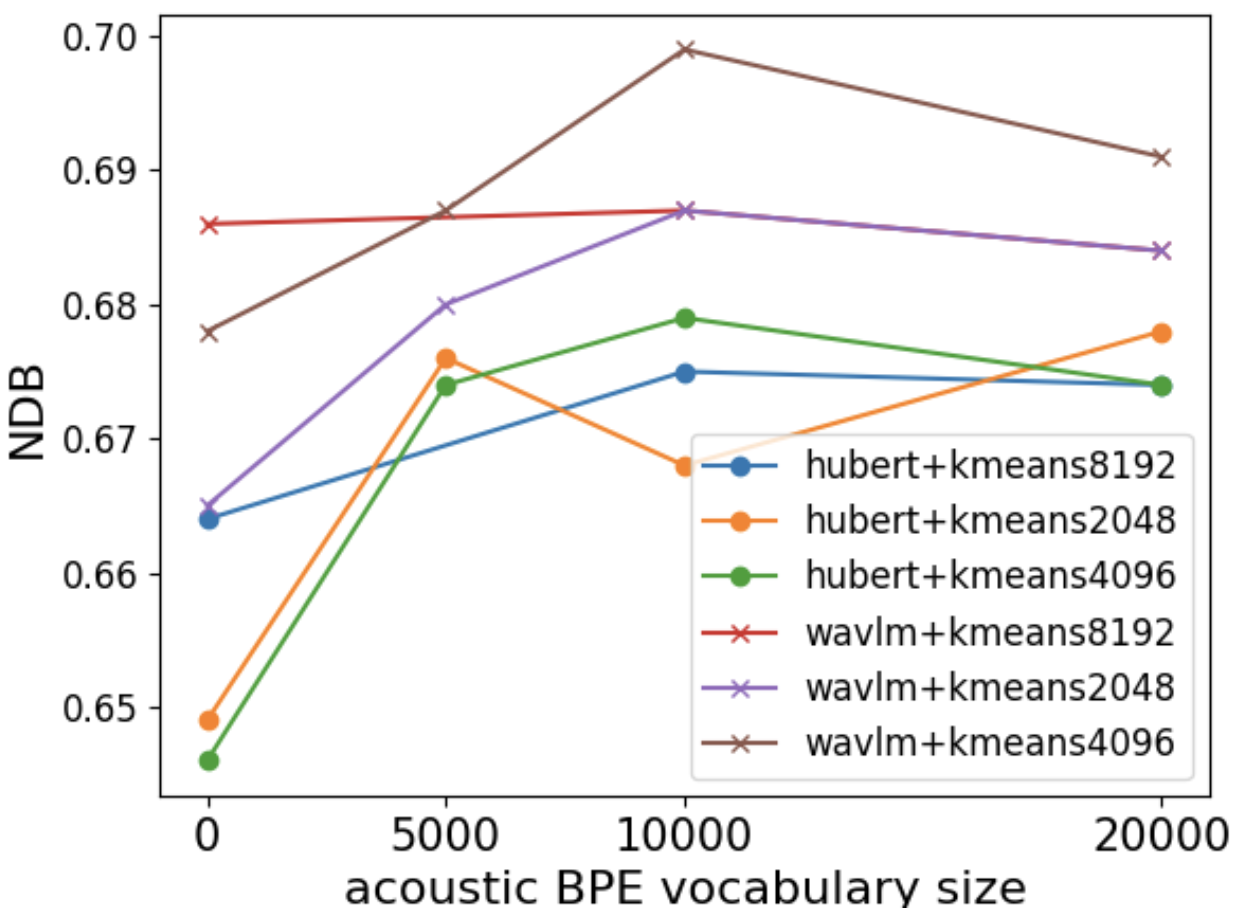}}
    \subfigure[JS]{\includegraphics[width=1.5in]{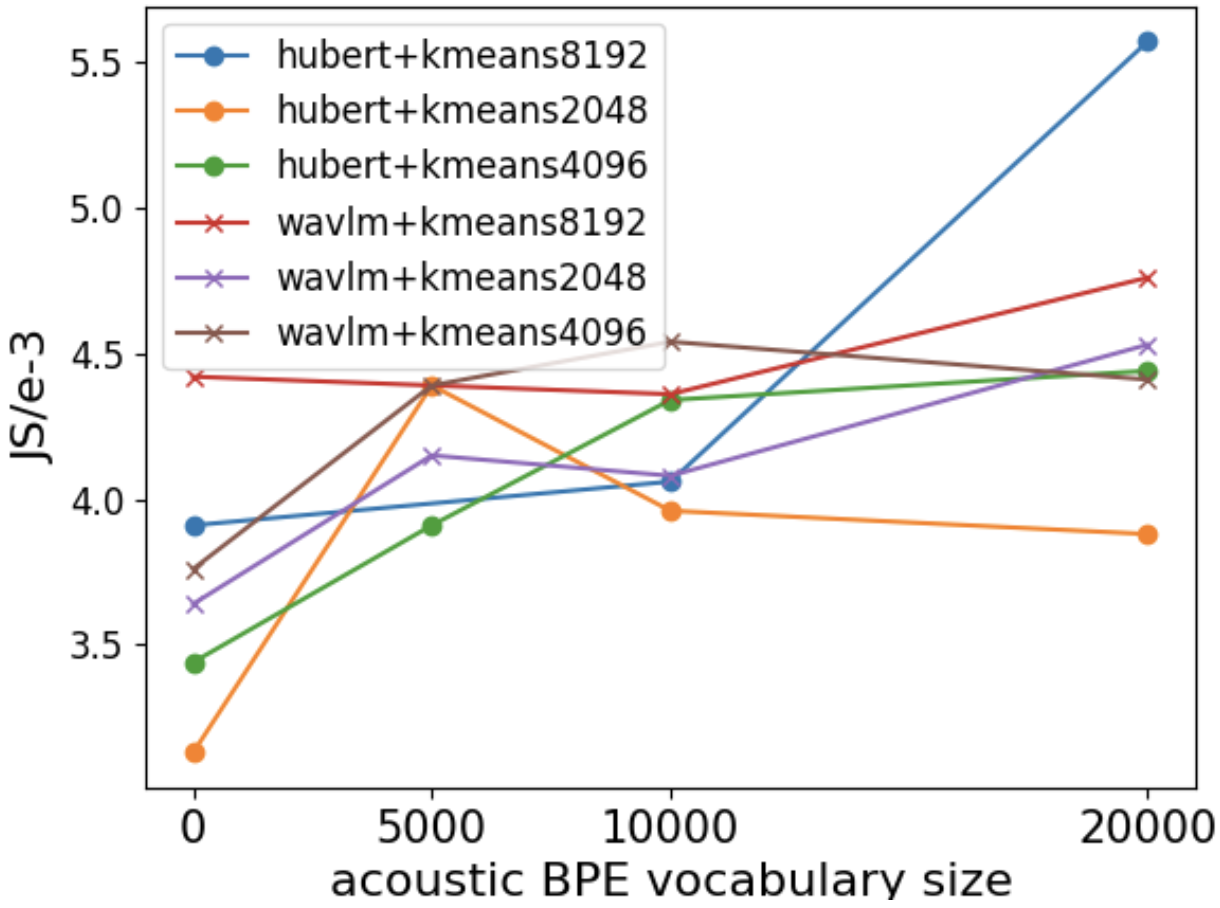}}
    \caption{NDB and JS divergence results under varied semantic token and acoustic BPE settings.}
    \label{figure1}
    \vspace{-0.5cm}
\end{figure}

\subsubsection{Increment of TTS sample diversity with acoustic BPE}
According to the presented results,  TTS models employing the acoustic BPE method showed significant advantages in synthesized audio diversity compared to those without it. Referring to Figure \ref{figure1}, this phenomenon is essentially prevalent. We will discuss extreme boundary cases in detail in next section \ref{discussion}.

\subsection{Discussion on boundary cases and limitations}
\label{discussion}
Although the benefits of the acoustic BPE in improving model performance are substantial, every method has its applicable range and limitations. 
This is particularly observed in our study, where there are configurable parameters that can be flexibly adjusted. 
Exploring the boundary conditions of this method under our research conditions can provide intuitive insights for constructing acoustic BPE methods in other decoder-only LM-based TTS systems with complex configurations. 
According to the results in Table \ref{Table3}, both too small and too large number of k-means centroids can lead to a plateau or even a decline in model performance. 
In practical experiments, when there is a significant gap between the number of k-means centroids and vocabulary size, instability may arise, e.g. leading to the model repeatedly outputting the same token. 
Moreover, WavLM also exhibits instability in the TTS architecture used in this experiment. 
At this time, the use of acoustic BPE method may exacerbate this instability, resulting in worse TTS performance.

\section{Conclusion}
In conclusion, the application of the acoustic BPE method in TTS tasks brings significant performance benefits, including improvements in the intelligibility, quality, and diversity of generated audio, as well as notable enhancements in training and inference speed. This undoubtedly demonstrates the potential of the method in discrete speech language modeling and speech-text language modeling. However, the presence of the acoustic BPE vocabulary increases the number of discrete tokens in speech. This imposes certain limitations on its configuration selection. In future work, we will conduct experiments with this method under scaled up datasets and models. Additionally, we will explore other effective methods for tokenizing audio.

\section{Acknowledgements}
This work was supported by the China NSFC Project (No. 92370206), Shanghai Municipal Science and Technology Major Project (2021SHZDZX0102) and Development Program of Jiangsu Province, China (No.BE2022059).


\bibliographystyle{IEEEtran}
\bibliography{paper}

\end{document}